\documentclass[10pt, conference, compsocconf]{IEEEtran}
\ifCLASSINFOpdf
\else
\fi
\usepackage{graphicx}
\usepackage{url} 

\begin{document}
%
\title{Out of vocabulary words decrease, running texts prevail and hashtags coalesce: Twitter as an evolving sociolinguistic system}




%
\author{\IEEEauthorblockN{Suman Kalyan Maity, Bhadreswar Ghuku, Abhishek Upmanyu and
Animesh Mukherjee}
\IEEEauthorblockA{Department of Computer Science \& Engineering\\
Indian Institute of Technology Kharagpur, India-721302\\ Email: \{sumankalyan.maity, bhadreswar.ghuku, abhishekd, animeshm\}@cse.iitkgp.ernet.in}}


\maketitle

\begin{abstract}
Twitter is one of the most popular social media. Due to the ease of availability of data, Twitter is used significantly for research purposes. Twitter is known to evolve in many aspects from what it was at its birth; nevertheless, how it evolved its own linguistic style is still relatively unknown. In this paper, we study the evolution of various sociolinguistic aspects of Twitter over large time scales. To the best of our knowledge, this is the first comprehensive study on the evolution of such aspects of this OSN. We performed quantitative analysis both on the word level as well as on the hashtags since it is perhaps one of the most important linguistic units of this social media. We studied the (in)formality aspects of the linguistic styles in Twitter and find that it is neither fully formal nor completely informal; while on one hand, we observe that Out-Of-Vocabulary words are decreasing over time (pointing to a formal style), on the other hand it is quite evident that whitespace usage is getting reduced 
with a huge prevalence of running texts (pointing to an informal style). We also analyze and propose quantitative reasons for repetition and coalescing of hashtags in Twitter. We believe that such phenomena may be strongly tied to different evolutionary aspects of human languages. 
\end{abstract}


%
\IEEEpeerreviewmaketitle

\section{Introduction}

Owing to rapid growth and penetration of the Internet in 21st century, the online social networks (OSNs) have become a de facto standard for sharing information, thoughts, ideas, personal feelings, daily happenings etc. The massive popularity of these sites has made available a huge amount of data of user interactions that offers unprecedented opportunities for analyzing and examining the data to infer how human society functions and evolves linguistically at scale. Therefore, a considerable fraction of research communities have shifted their focus to these sites. Among the OSNs, Twitter has emerged as the most popular medium of research due to ease in access of content and less privacy constraints. A huge volume of research works spanning various fields like computer science, social science, physics etc. has been done on Twitter data. To name a few recent studies, one can highlight the works on retweeting behavior~\cite{mac,rec,boyd}, deleting tweets~\cite{alm,petro}, trending topics~\cite{
yang,huang}, popularity of hashtags and its spreading~\cite{bigbird,viral,game} etc.

Twitter houses many features that makes its language very distinct from other social media. Due to the hard 140-character limit in tweets, its language is very brief and compact and hence comparable to SMS and online chats; however Twitter also provides opportunity for discussion of a much wide variety of topics from daily chitchatting to news events, sports gossip and some serious discussions which are not part of SMS and online chats. A recent study by Hu et al.~\cite{dude} has tried to analyze the linguistic differences across various media and claimed that Twitter surprisingly possess more formal linguistic traits than it has been believed to.

\subsection*{Motivation}
Though the Twitter research community is growing rapidly, there has been relatively very little work that has tried to observe the behavioral, linguistic, psychological aspects prevalent in the tweets at larger scale. Most of the studies mentioned earlier had focused on various problems on Twitter but on shorter timescales ranging from few weeks to few months only. An equivalent study on larger timescales is very important because it can reveal many long-term trends that would remain otherwise unobserved. A very recent study by Liu et al.~\cite{mislove} has attempted to study the users' activity and their behavior on a very large time scale from 2006 to 2013. This study has showed how some of the previous research outcome gets invalidated; for example, the $32\%$ retweet fraction mentioned in ~\cite{mac} is found to be only $10\%$ by them. This motivates us to study Twitter as an evolving sociolinguistic system at a larger time scale.

For developing any application on any enterprise media or development of cognitive assistants/mediators for smart service systems, one needs to build a strong analytic handle. A major part of this is language and text analytics. Our study could help improving the cognitive assistants for smart service systems like IBM Message Resonance\footnote{http://www.ibm.com/smarterplanet/us/en/ibmwatson/developercloud/message-resonance.html}, IBM Tone Analyzer\footnote{ \url{http://www.ibm.com/smarterplanet/us/en/ibmwatson/developercloud/tone-analyzer.html}}, Textio\footnote{\url{https://textio.com/}} etc. We have studied formality/informality aspects of Twitter texts (see section 5), the ranking of the words and hashtags based on core-periphery analysis (see section 6 and 8) and hashtag compounding. The cognitive assistants can be informed by these above linguistic styles - formality, compounding etc., to improve their performance. IBM Message Resonance is a cognitive assistant that communicate with people with a style and words that suits them. Formality/informality is also a linguistic style that can be incorporated with it for better quality and customer service. Similarly, IBM Tone Analyzer is a cognitive assistant that analyzes linguistic tones of writing which can also be improved with the formality feature. Apart from this, most of these above cognitive assistants are built on the words that are in-vocabulary; however social media text include many out-of-vocabulary words and analysis involving them shall certainly boost performance of the systems. Further, our methodology of the temporal core-periphery analysis of the words and hashtags can be useful for smart services like in the identification and recommendation of popular hashtags and also beneficial for trend identification.
\subsection*{Organization of the paper}
In this paper, we have dissected the Twitter media to bring out various 
linguistic and sociolinguistic aspects and their evolution over time. The remainder of the paper is organized as follows. Section 2 discusses about the state-of-the-art literature. In section 3, we describe the dataset briefly. Section 4 presents an analysis of the evolution of basic linguistic quantities. In Section 5, we attempt to investigate the (in)formality of Twitter texts. Section 6 presents the network level analysis of the Twitter words. In section 7, we discuss about evolution of various entities in Twitter. Quantities like number of mentions show an overall increase indicating that Twitter is increasingly becoming a conversational media. In section 8, we perform a comprehensive study on the Twitter hashtags.  Some of the striking observations are i) that there are large fractions of tweets which exhibit repetition of a single hashtag (ii) hashtags frequently coalesce to form new hashtags and (iii) we can find certain cases where the popularity of the 
coalesced hashtags are orders of magnitude higher than the constituent hashtags; in all such cases, the overlap between the word cloud surrounding the constituent pair of hashtags forming the coalesce is found to be very high compared to a random pair of hashtags. Finally, in section 9, we conclude this study by summarizing our findings and outlining important future directions.

\section{Related Works}
\subsection*{Cognitive and linguistic studies in CMC} There have been various works on analyzing linguistic style, structure of language, as well as its cognitive aspects. Some early works includes the analysis of the cognitive process involved in picking words and the linguistic style~\cite{flower}, the variations across different registers~\cite{biber}, and the correlation between style and gender~\cite{carroll} by Carroll et al. With the advent of Internet, the research focus shifted towards the language of computer-mediated-communication (CMC) systems like online chats, IM etc. Paolillo in ~\cite{pao} investigate linguistic variations associated with strong and weak ties in an early Internet chat relay system. Thurlow et al. study the linguistic styles in SMS~\cite{thurlow}. Similar research of understanding linguistic styles have been carried out subsequently in various other media: Tagliamonte et al. in~\cite{tag} study the IM media, emails and blogs have been studied by Baron et al.~\cite{baron} and Herring et al.~\cite{herring} respectively. There have also been some studies on analysis of linguistic content and structure of deceptive CMC interaction and the linguistic profile of the sender and receiver~\cite{madhu,hancock}.

\subsection*{Content and linguistic analysis of the Twitter data}
The increasing popularity of Twitter media and the ease of accessing user data has also propelled a good amount of research works that primarily focus on the content and linguistic 
analysis of the Twitter data. An interesting linguistic activity in Twitter is the user interactions or the conversations. Java et al.~\cite{java} find that $21\%$ of Twitter users use this media for conversations and $12.5\%$ of all tweets are part of conversations. Similar investigations have been carried by Naaman et al.~\cite{naaman}. Honeycutt and Herring~\cite{hohe} analyze conversational exchanges in Twitter focusing on mentions. They find that short dyadic conversations occur frequently, along with some longer multi-participant conversations. Boyd et al.~\cite{boyd} study various conventions of retweeting practices in Twitter. Ruth Page~\cite{page} studies the contrasting ways in which corporations, celebrities and ordinary Twitter users use hashtags as a resource to seek attention of the mass with self-branding, self-promotions. Houghton and Joinson~\cite{join} identifies linguistic markers for self-disclosures and sensitive information in tweets.  Eisenstein et al. in~\cite{eisen} study the role of geography and demographics on the language in Twitter. Hong et al. in~\cite{hong} investigate the cultural differences in Twitter's language. Gruzd et al.~\cite{gruzd} study how happiness spread in Twitter. Nambisan et al.~\cite{namb} study the depressive disorder of Twitter users from their posts. Wamba and Carter~\cite{wamba} studies the impact of various factors in adoption of Twitter in various organizations.
\subsubsection*{Smart cognitive/linguistic systems using Twitter}
There have been several studies that use Twitter data to build smart cognitive and linguistic systems. Ramage et
al.~\cite{ram} develop a partially supervised learning model (Labeled LDA) to summarize key linguistic trends and features on a corpus of 8M Twitter posts. They identify four general types of dimensions: substance, status, social and style. These include dimensions about events, ideas, things, or people (substance), related to social communication (social), related to personal updates (status), and indicative of broader trends of language use (style). They use this summarization for personalized recommendation to Twitter users for finding people suitable to follow. Ritter et al.~\cite{alan} develop an unsupervised learning approach to identify conversational structure from open-topic conversations. They train an LDA model on a combined dataset of conversational (speech acts) and content topics of 1.3 million Twitter conversations, and identify interpretable speech acts (reference broadcast, status, question, reaction, comment, etc.) by clustering the similar conversational roles. Danescu-Niculescu-Mizil et al. in~\cite{dan} build on this data set in~\cite{alan} and extend it to include the complete conversational history of individuals over a period of almost one year. They study how people adopt linguistic styles while in conversation on Twitter. Duan et al. in~\cite{duan} employs a learning approach to rank tweets considering both twitter specific features in conjunction to textual content. Apart from this, various other machine learning-based approaches have been proposed to enhance the textual features of Twitter language~\cite{ritter,owu,hua}.

\section{Dataset description}
Twitter provides $1\%$ random sample of all the tweets via its sample API in real time. We use this API to crawl tweets from $1^{st}$ January, 2012 to $31^{st}$ December, 2013. For analysis, we consider the users who have mentioned English as their language in their profile. We also performed a second level filtering of the tweets by a language detection software~\cite{langid} to remove any non-English tweets in our dataset. We then tokenize and POS tag by CMU POS tagger~\cite{owu} which is the state-of-the-art tagger for Twitter data. In total the dataset consists of $\sim$1 billion tweets.

\section{Evolution of the basic linguistic quantities in Twitter}
\begin{figure}[h]
\begin{center}
\includegraphics*[width=1\columnwidth, angle=0]{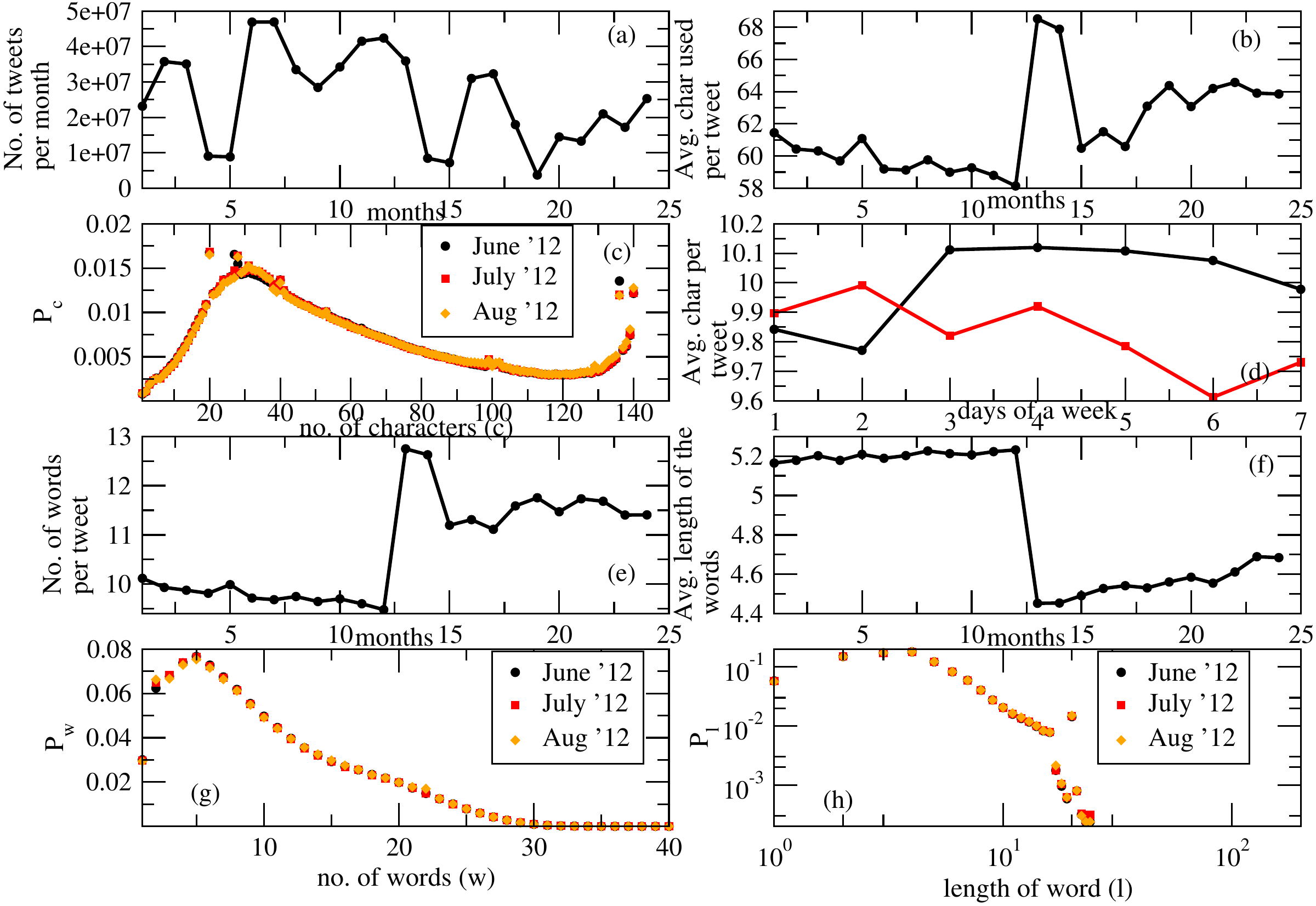}
\caption{\label{fig1} Evolution of basic linguistic quantities a) No. of tweets over the months (1 in x-axis refers to Jan 2012 and 24 refers to Dec 2013) b) monthwise avg. character usage per tweet c) Distribution of character usage over three representative months d) daywise avg. char usage in two random weeks of the whole data (1-7 refers to Monday to Sunday) e) No. of word usage per tweet monthwise f) evolution of avg. word length over the months g) Distribution of no. of words used in tweets over three representative months h) Distribution of word lengths over the tweets.}
\end{center}
\end{figure}

Since its inception in 2006, Twitter has grown rapidly. From a small base of users in 2006, it has close to billion users in 2013. With its rapid growth, the usage of Twitter as a media has significantly changed. More organizations, individuals are using Twitter as a primary source of information dissemination. Not only the sociological aspects, Twitter is also evolving as a linguistic system. In this section, we study the evolution of some of the basic linguistic characteristics of Twitter. In fig~\ref{fig1}(a), we show how the no. of tweets is changing over time. As time progresses, we observe that usage of characters per tweets is increasing. Fig~\ref{fig1}(b) clearly shows that the avg. character usage per tweet has sharply risen from 2012 to 2013. To dig deeper into the character usage, we plot the probability distribution of the character usage within the tweets. We find two distinct peak in the distribution - one at $\sim 30$ characters and other at $140$ characters. Therefore, though 
avg. character usage is $\sim 61.8$, people do utilize the whole of 140 character-limit quite frequently. The distribution is consistent across different months (see fig~\ref{fig1}(c)). Fig~\ref{fig1}(d) shows the weekwise average character usage that points to higher usage of characters in the middle of the week than the start and the weekend.

Next we present a few observations related to the immediate higher level linguistic unit, i.e., words. As time progresses, no. of words per tweets are increasing (fig~\ref{fig1}(e)) whereas probably due to the 140-character hard limit, the average length of the words are decreasing (fig~\ref{fig1}(f)). Therefore, in Twitter people are using more short forms to communicate among themselves. For the completeness of the word usage study, we also plot the distribution of word usage in the tweets within a month. The fig~\ref{fig1}(g) shows a peak at $\sim 5$ words whereas from the distribution of the word length in fig~\ref{fig1}(h), it is evident that 2-4 length words are used more.
\section{Is Twitter Informal?}
Twitter is known to be more informal and close to SMS/online chat language. However, a recent study by Lu et al.~\cite{dude} claim that Twitter is markedly more standard and formal than SMS and online chat and more close to email and blogs. The reasons they point out do not clearly suggest/deny their claims. They also suggest that Twitter's language is not too extreme in uniqueness so that one can claim Twitter to be a departure from English language\footnote{http://www.telegraph.co.uk/culture/film/8853427/Ralph-Fiennes-blames-Twitter-for-eroding-language.html}. We attempt to understand the formality vs informality issue through the introduction of new metrices here. There is a trend in SMS, online chats etc of using running texts with no space between two or more words. This is a standard notion of informalism. 
\begin{figure}[h]
\begin{center}
\includegraphics*[width=1\columnwidth,angle=0]{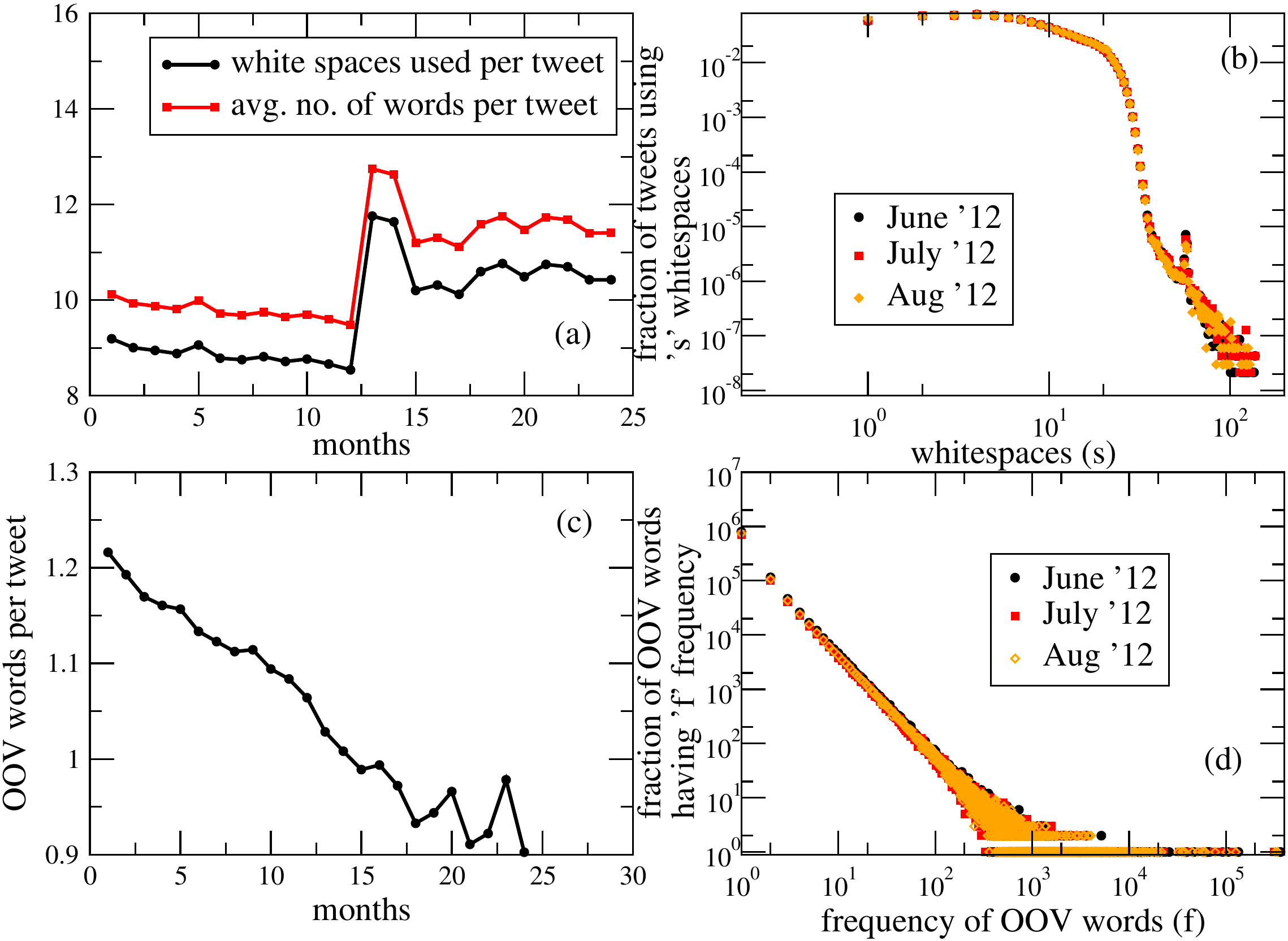}
\caption{\label{fig2} a) Monthwise usage of whitespaces per tweet compared to no. of words used per tweet b) Distribution of whitespace used in tweets for sample 3 months data c) Monthwise occurrences of Out-of-Vocabulary (OOV) words d) Distribution of frequency of the OOV words in the tweets.}
\end{center}
\end{figure}
In fig~\ref{fig2}(a), we show how the use of whitespaces is evolving over the months. Note that ideally the number of white spaces should be an indicator of the number of words present in a tweet. However, the number of words obtained as the output of the tokenizer is clearly much higher than the number of whitespaces  according to fig~\ref{fig2}(a). This implies a strong evidence of the presence of running text and the trend is persistent throughout the two year timeline. We also find the distribution of the whitespace usage in the tweets within months which is clearly pointing to high usage of 2-5 whitespaces (see fig~\ref{fig2}(b)). Therefore, it might be tempting to identify the linguistic style of Twitter to be informal. However, the analysis of Out-of-Vocabulary (OOV) words across tweets portray a different picture. To this purpose, we use the PyEnchant dictionary which is Python's spellchecking dictionary\footnote{https://github.com/rfk/pyenchant} to identify the words which are not present in the 
standard English library. Figure~\ref{fig2}(c) shows the monthwise usage of OOV words per tweet. The graph shows a decreasing trend which invalidates our
notion of informalism established through whitespace usage characteristics. The results together possibly points to the fact that Twitter is nor completely informal like SMS/chat languages, neither is it as formal as the standard English language. In contrast, it seems to be somewhere in between these two extremes.

\section{Network level study of the words in Twitter}
\begin{figure}[h]
\begin{center}
\includegraphics*[width=1\columnwidth,angle=0]{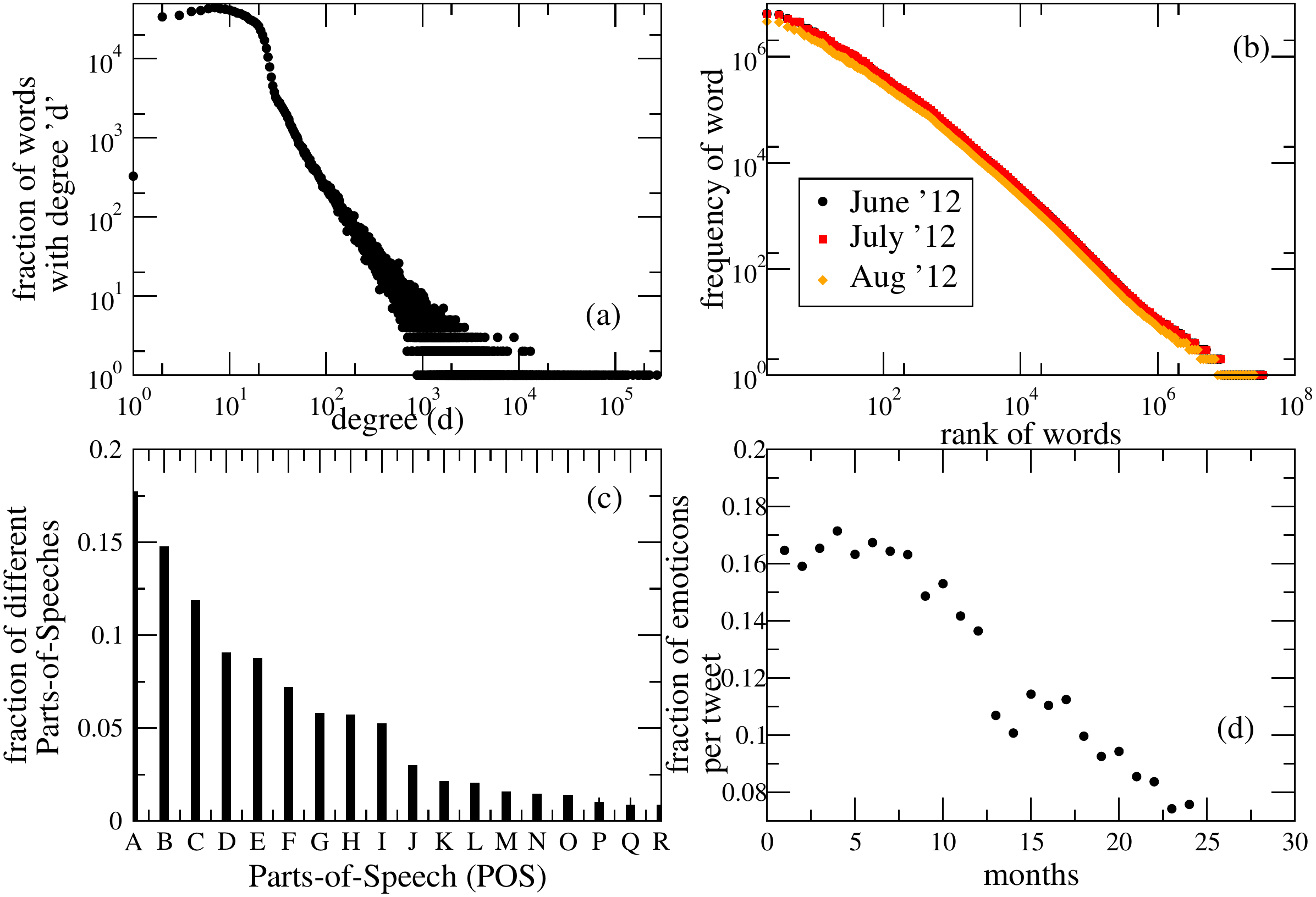}
\caption{\label{fig3} a) Degree distribution of word co-occurrence graph b) Zipf's law c) Distribution of various POSes over a month data. Different symbols A, B, C, D, ..., R refers to verb, common noun, punctuation, pre or postposition, pronoun, determiner, adverb, adjective, proper noun, interjection, nominal + verbal, Coordinating conjunction, Emoticons, URLs, numeral, discourse marker (indicating continuation of message across multiple tweets), symbols and abbreviations and foreign words, verb particle respectively d) Monthwise distribution of emoticons per tweet.}
\end{center}
\end{figure}

In this section, we study the interactions between the linguistic units of Twitter language. We construct the word co-occurrence graph by forming edges between the words that co-occur in the same tweet. In fig~\ref{fig3}(a), we show the degree distribution of the word co-occurrence graph. This distribution has two parts - the first part is like a poisson distribution while the next part is a power-law distribution with a heavy tail. If we closely look into the words in different parts of the distribution, we find that the random part of the distribution mostly contains mentions, URLs and some nonsense words. For example, words like eeeeeeeeeeeeee, @franzyy10, @im\_not\_sarboat, @cindy\_seororo22, @krisneil49kling, http://t.co/52dqqpmr, http://t.co/5jo7gcvd, zombieism. These tweets seem to originate from more general users. Now if we analyze the torso portion of the distribution of the graph, we find that those words also contain mentions but which are more popular like @cristiano, @drunkbarney, @nashville unlike in the random portion. The other words in this region are also more meaningful compared to the words in random part. The words in the tail region of the power law graph are the high degree words. For example, emoticons and slangs like :) , :(, LOL, OMG and general Twitter words like i'm, that etc. dominate this region.

Zipf's law states that given some corpus of natural language utterances, the frequency of any word is inversely proportional to its rank in the frequency table. Thus the most frequent word will occur approximately twice as often as the second most frequent word, three times as often as the third most frequent word and so on. Here, we attempt to verify whether Twitter's language follows Zipf's law. Fig.~\ref{fig3}(b) shows the frequency vs rank plot which follows the general trend of Zipf's law. We use the CMU POS tagger~\cite{owu} to tag various words in the Twitter corpus. In fig~\ref{fig3}(c), we show the occurrence of various Parts-of-Speech (POS) tags per tweet. The most used POS is verb, followed by common nouns. This rank order of frequency distribution of the POS tags is consistent across all other months, however the contribution of all the POSes in the whole dataset does not remain same. In fig~\ref{fig3}(d), we show the evolution of emoticons over Twitter. It seems that the usage of emoticons per 
tweet has decreased over the years.

\begin{figure}[h]
\begin{center}
\includegraphics*[scale = 0.3,angle=0]{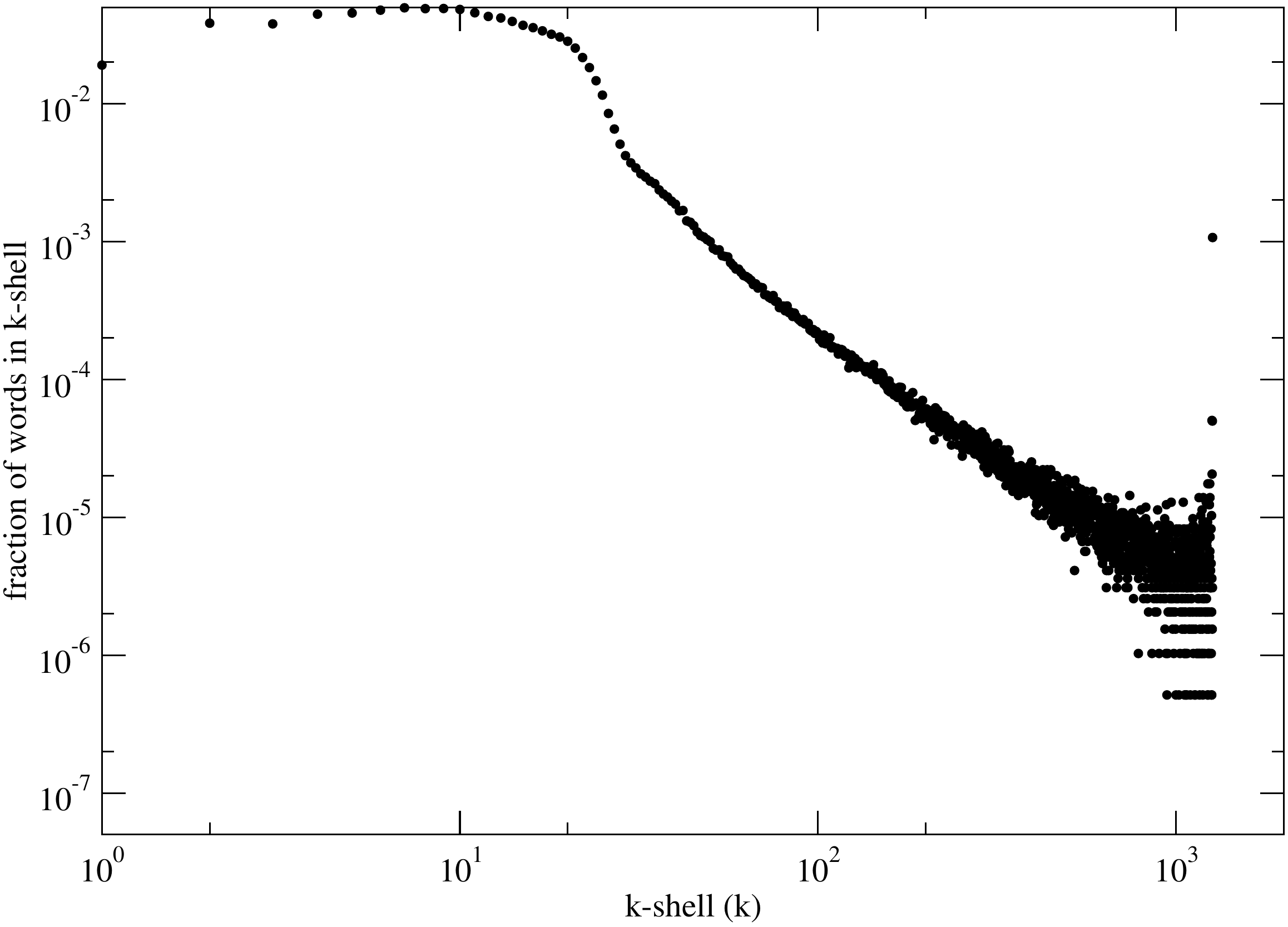}
\caption{\label{fig4} K-shell distribution of the words for a representative month. }
\end{center}
\end{figure}

We further analyze the word co-occurrence graph to have better understanding of the network structure. To identify important nodes in a graph, degree could be misleading. Therefore, we resort to K-shell decomposition method to find underlying hierarchies of importance in the graph. In fig~\ref{fig4}, we show the distribution of the words across various shells. The distribution is dominated by a large power-law part and a small random part in the beginning. It turns out that the innermost shell ($k_{max}$) has a large fraction of words in it clearly suggesting the existence of a core in the network.

\begin{figure}[h]
\begin{center}
\includegraphics*[width=1\columnwidth,angle=0]{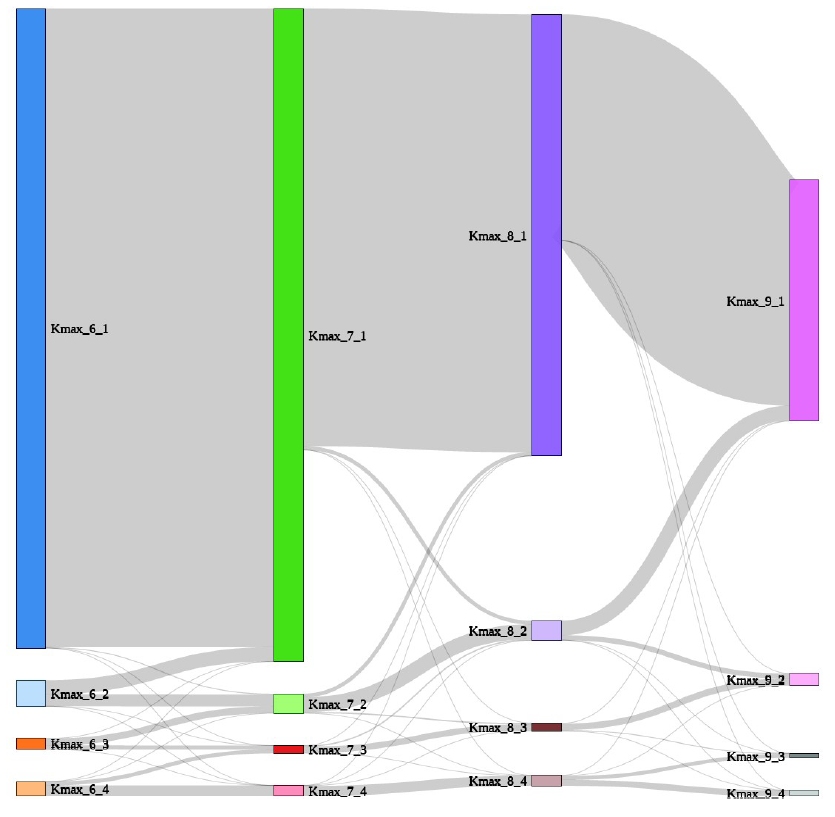}
\caption{\label{fig5} Migration of words from various region of k-shells over 4 consecutive months.}
\end{center}
\end{figure}

For analyzing the stability of this core-periphery structure, we perform the migration analysis of various words across shells. We divide the words into four categories based on their K-shell indices by dividing the range of K-shell values into four groups of approximately equal sizes. Thus Region I contains words that are in the core of the network ($k \in [\frac{3}{4}k_{max} , k_{max}]$), and Regions II, III, and IV contain nodes with increasingly lower K-shell indices. Fig~\ref{fig5} shows the migration of various words across 4 regions for four consecutive months data from June 2012 to September 2012. There are very few words which go from the core regions to peripheral regions. This suggests that words in the higher shells remain stable over period of time. Migration usually takes place where there is a significant event propelling  some words related to the event to change their current shells. Some examples of shell migration are as follows ``@darrencris'' migrated from shell number 9 of July 2012 
to shell number 684 of August 2012, which is a huge increase suggesting that it became popular . ``\#100thingsaboutme'' migrated from shell number 911 to shell number 4, which indicates a decrease in popularity of this hashtag.

\section{Evolution of entities in Twitter}
\begin{figure}[h]
\begin{center}
\includegraphics*[width=1\columnwidth,angle=0]{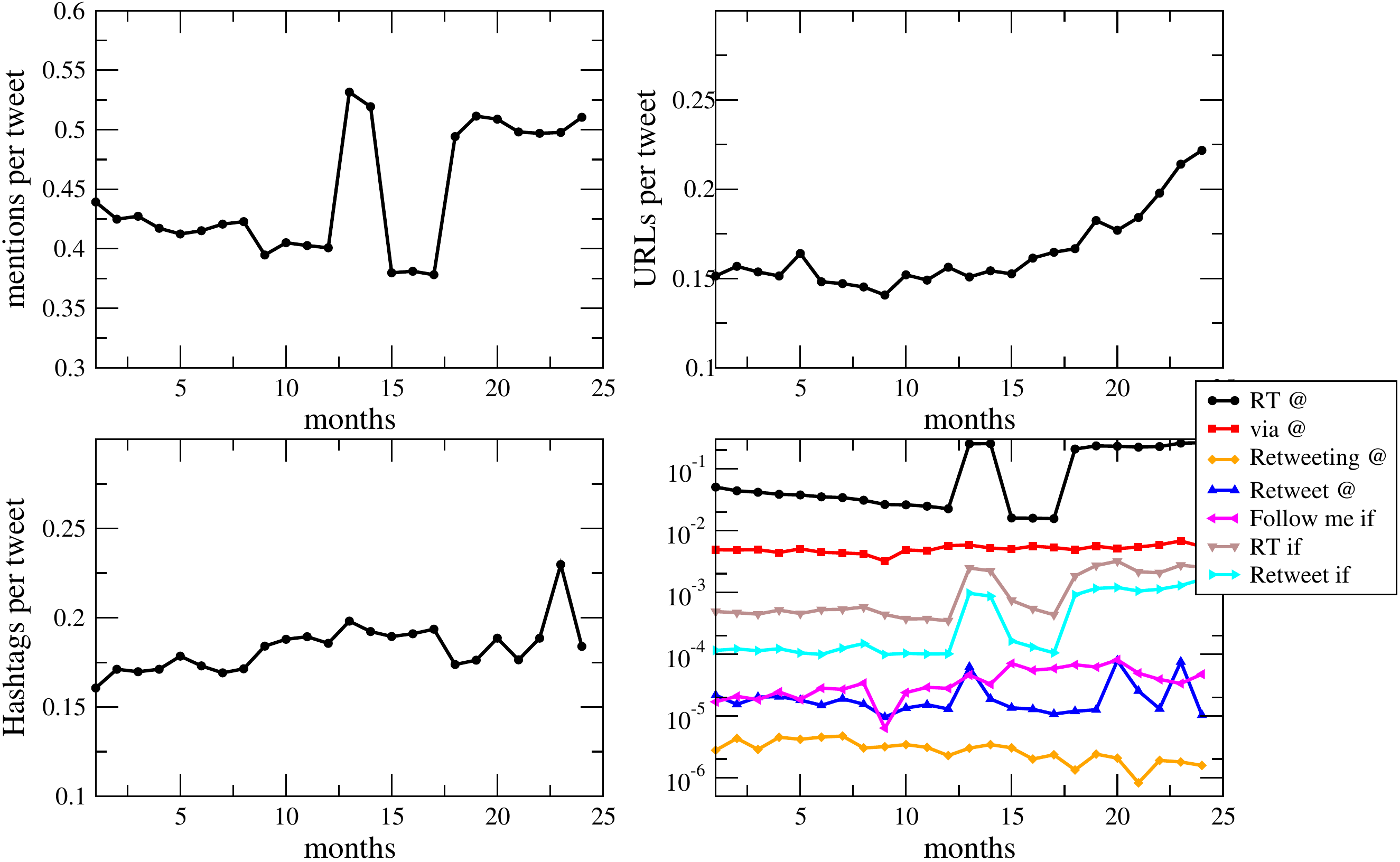}
\caption{\label{fig6} Evolution of various entities in Twitter over the time period of 24 months a) No. of mentions per tweet b) No. of URLs per tweet c) No. of hashtags per tweet d) various Retweeting conventions per tweet.}
\end{center}
\end{figure}

Twitter allows its users to use various entities. One can use mentions to tag their friends, celebrity pages etc. People use hashtags to organize, categorize, find conversations on various topics. Twitter also provides a mechanism to share information through tiny URLs that are extensively used by a large number of Twitter users. There is another unique feature that Twitter possesses in the form of retweeting conventions. In Twitter, there are many variety of Retweeting conventions are used, though RT and via are the most popular ones. In this section, we shall study these entities and their usage in the tweets over the years. Fig~\ref{fig6}(a) shows the evolution of mentions per tweets. About 50\% of the tweets contain a mention which further has increasing trend clearly suggesting that Twitter is now more often used for direct conversations. Similarly, the use of hashtags and URLs are also increasing (see fig~\ref{fig6}(b) and (c) respectively). RTs are mostly used in Twitter followed by via. The other 
conventions (though not so popular) are also used in Twitter in non-negligible fractions. The existence of these different retweeting conventions suggest that people are quite selective about adopting the linguistic styles.

\section{Hashtag as a new paralanguage}
Hashtag is the new ``paralanguage'' of Twitter. What started as a way for people to connect with others and to
organize similar tweets together, propagate ideas, promote specific people or topics has now grown into a language of its own. As hashtags are created by people on their own, any new event or topic can be referred to by a variety of hashtags. This linguistic innovation in the form of hashtags in one very special feature of Twitter which has become hugely popular and are also used in various other social media like Facebook, Google+ etc. and have been studied extensively by researchers to analyze the competition dynamics, the adoption rate and popularity of these hashtags. However, there are very few attempts to study the linguistic aspects of hashtag evolution over large time scales.
Thus, it is interesting and worthwhile to analyze the evolution of the usage of hashtags and their linguistic aspects. Fig~\ref{fig6} shows that there is more or less an increasing trend in hashtag usage over the years and $\sim 17-20 \%$ of the tweets are having one or more hashtags. We devote this section to analyze in detail the hashtags as a special linguistic unit of Twitter.

\begin{figure}[h]
\begin{center}
\includegraphics*[width=1\columnwidth,angle=0]{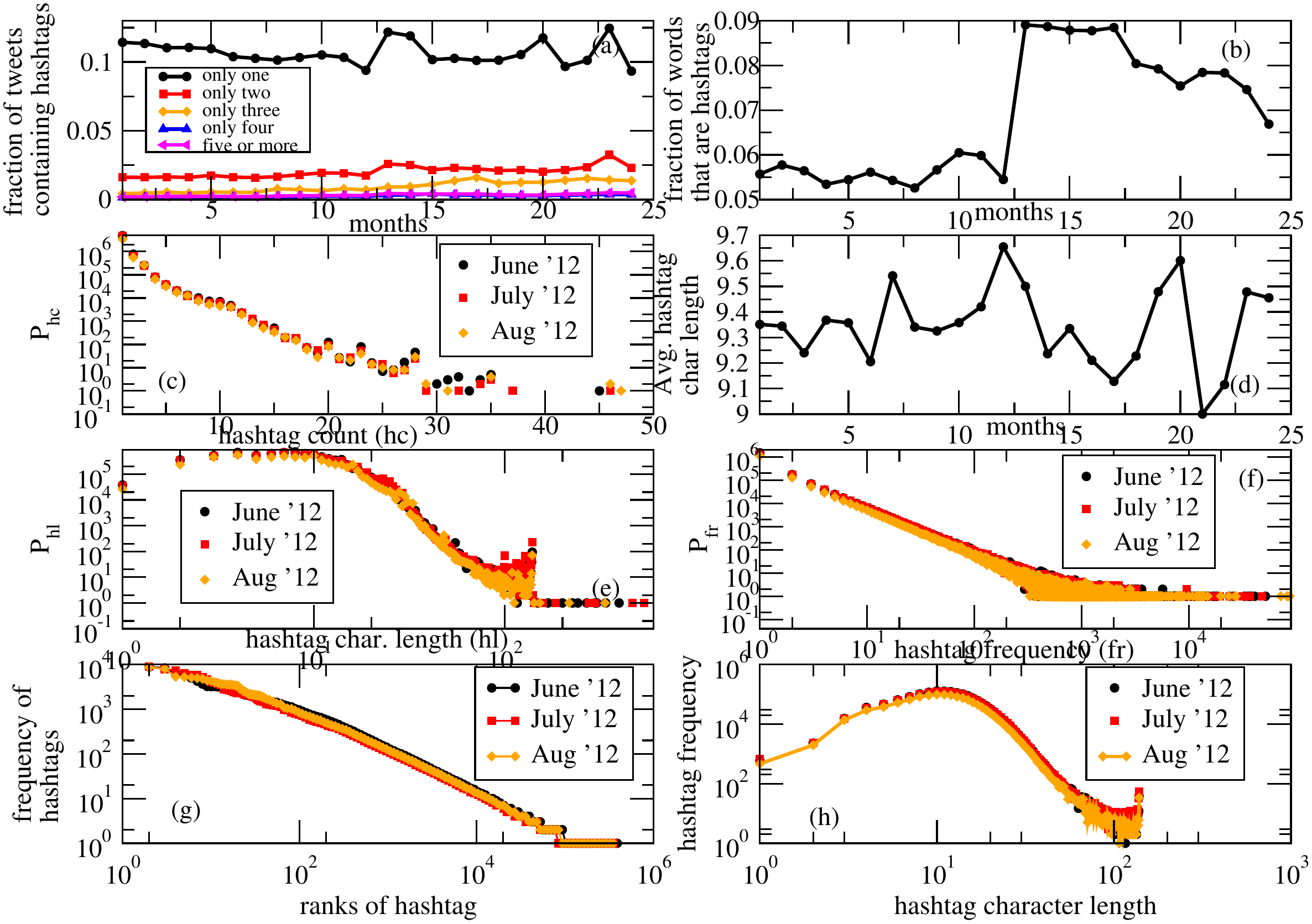}
\caption{\label{fig7} a) Monthwise distribution of number of tweets containing only one hashtag, only two hashtags, three hashtags, four hashtags, five and more than five hashtags b) monthwise distribution of fraction of words that are hashtags c) Distribution of count of hashtags in tweets over three representative months d) Monthwise avg. usage of characters in hashtags e) Distribution of hashtag character lengths over three representative months f) Distribution of hashtag frequencies over three representative months g) plot of hashtag frequency vs rank of hashtags h) Graph of frequencies of hashtags vs character lengths}
\end{center}
\end{figure}

Fig~\ref{fig7} shows the evolution of basic linguistic features of hashtags. In fig~\ref{fig7}(a), we show how the hashtags are distributed across the tweets. About $12\%$ and $2.5\%$ of the tweets contains single hashtags and double hashtags respectively. For single hashtags, this fraction remains more or less stable as time progresses while the fraction of tweets containing multiple hashtags increases. This is another observation which motivates us to investigate why more than one hashtags are used in a tweet and we shall discuss them separately in a subsequent subsection. In fig~\ref{fig7} (b), we show the evolution of the contribution of hashtags in the total vocabulary of Twitter words. It is quite evident that the fraction of words that are hashtags are increasing over the years - $\sim 6\%$ in 2012 to $\sim 8\%$ in 2013. In fig~\ref{fig7}(c), we plot the distribution of no. of hashtags occurring in tweets. This distribution follows power-law with a non-negligible fraction of tweets containing 
more than 5-6 hashtags. Even in some cases, we found $\sim 30-40$ hashtags being used. Some of the examples of tweets containing a large no. of hashtags are presented here.
\begin{itemize}
\item \#me \#friend \#coco \#night \#nightclub \#black \#party \#fun \#two \#girls \#happy \#smile
\#eyes \#pic \#noche http://t.co/hcjQdSnU
\item Friday night. \#JESUSgirl \#keiramachae \#beautiful \#white \#sweater \#jeans \#heels
\item \#me \#flowery \#hat \#AKBstyle \#nice \#famztime \#holiday \#jtp2 \#batu \#pixlrekspres. :)
http://t.co/06QTnRVj
\item My first keek video! Yay :) \#country \#singer \#songwriter \#cat \#facebook
\#twitter \#youtube \#reverbnation \#itunes http://t.co/75jJ3qoA
\end{itemize}
These type of tweets are generally expressing strong feelings and excitements. Next, we observe the character usage in the hashtags which shows an average of $\sim 9.3$ (see fig~\ref{fig7}(d)). In fig~\ref{fig7}(e), the hashtag character length distribution has been shown which follows poisson distribution at the initial part and a power-law afterward with more than expected no. of hashtags of length $\sim 100$. We also found instances of complete tweets corresponding to one single hashtag. These tweets are created by merging couple of words together to possibly utilize the space restriction in Twitter better. For example, these are mostly Twitter idioms or babbles. We also find smaller length hashtags in the form of abbreviations, general slangs etc. Some of the examples are the following:
\#eh, \#DT, \#NF, \#eh, \#np, \#fb, \#JJ, \#nw, \#Nf, \#RT etc.

Though people create and use a lot of hashtags; not all of them become popular. Most of them are not used at large scale whereas some of them become highly popular. Fig~\ref{fig7}(f) shows the distribution of hashtag frequencies in various months. The distribution clearly indicates a power-law behavior with a heavy tail. We also observe the relationship between the frequency of the hashtags and their ranks in the corpus. The relationship follows zipf's law (see fig~\ref{fig7}(g)). There are many works that try to predict the popularity of hashtags~\cite{bigbird,viral,game}. One of the linguistic features for the popularity of the hashtags is indicated by the hashtag character length. To validate this hypothesis, we observe the relationship between hashtag frequency and their character length over various month data. The fig~\ref{fig7}(g) shows that there is no direct inverse relationship existing among them. It behaves more or less like normal distribution with hashtags having mean character length ($\sim$10) achieving highest frequency.

\begin{figure}[h]
\begin{center}
\includegraphics*[scale = 0.3,angle=0]{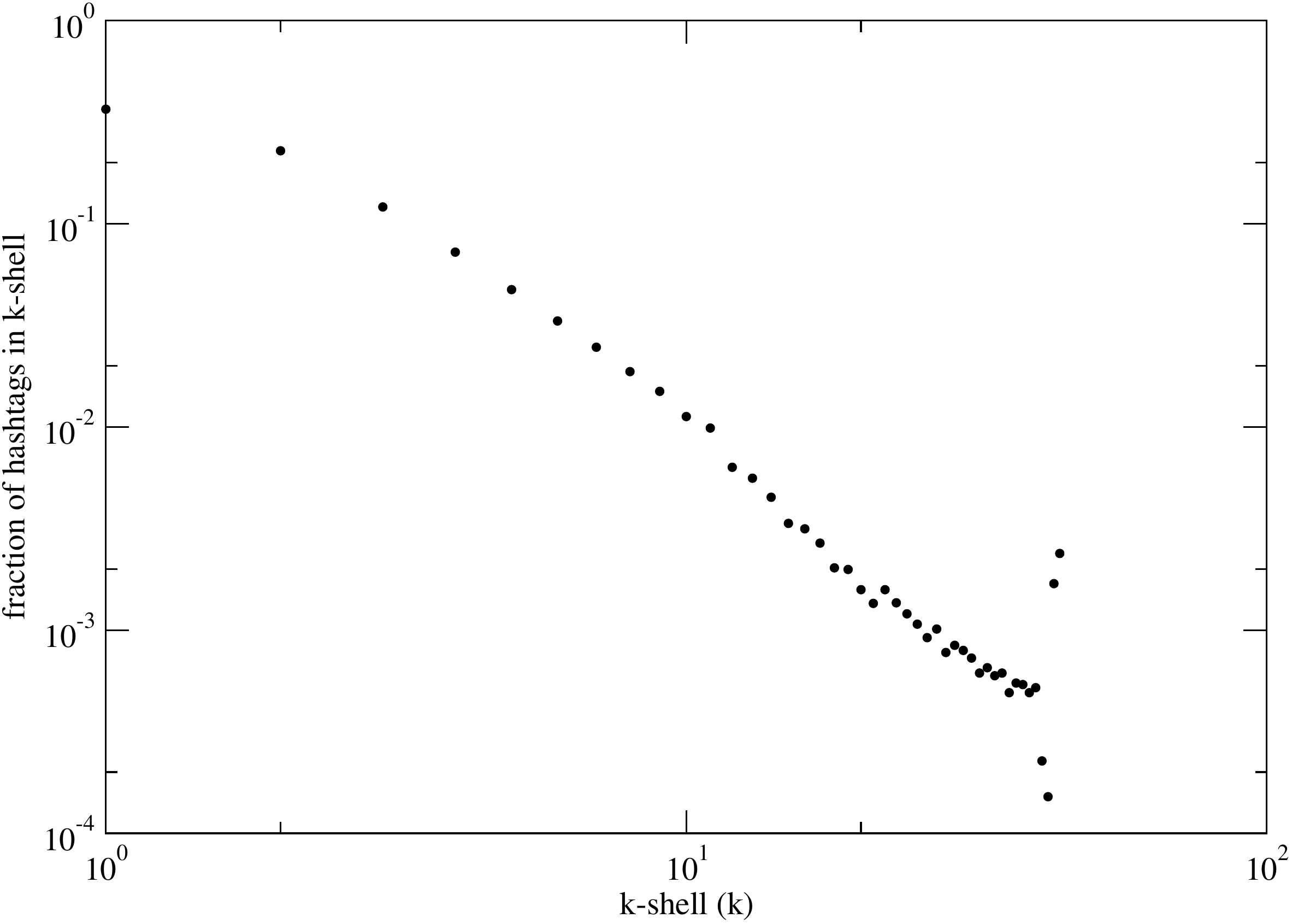}
\caption{\label{fig8} K-shell distribution of the hashtags for a representative month. }
\end{center}
\end{figure}

Similar to word co-occurrence networks, we also construct hashtag co-occurrence networks. To identify important hashtags in the hashtag co-occurrence graph, we adopt K-shell decomposition method to find underlying hierarchies of importance in the graph. In fig~\ref{fig8}, we show the distribution of the hashtags across various shells. The distribution is dominated by a small random part in the beginning followed by a large power-law part similar to what we observe in fig~\ref{fig4}. The innermost shell ($k_{max}$) has quite a large fraction of hashtags compared to most of the other shells in it clearly suggesting an existence of dense core in the network. The fraction of nodes in $k_{max}$ is also higher compared to the word counterpart.

\begin{figure}[h]
\begin{center}
\includegraphics*[width=1\columnwidth,angle=0]{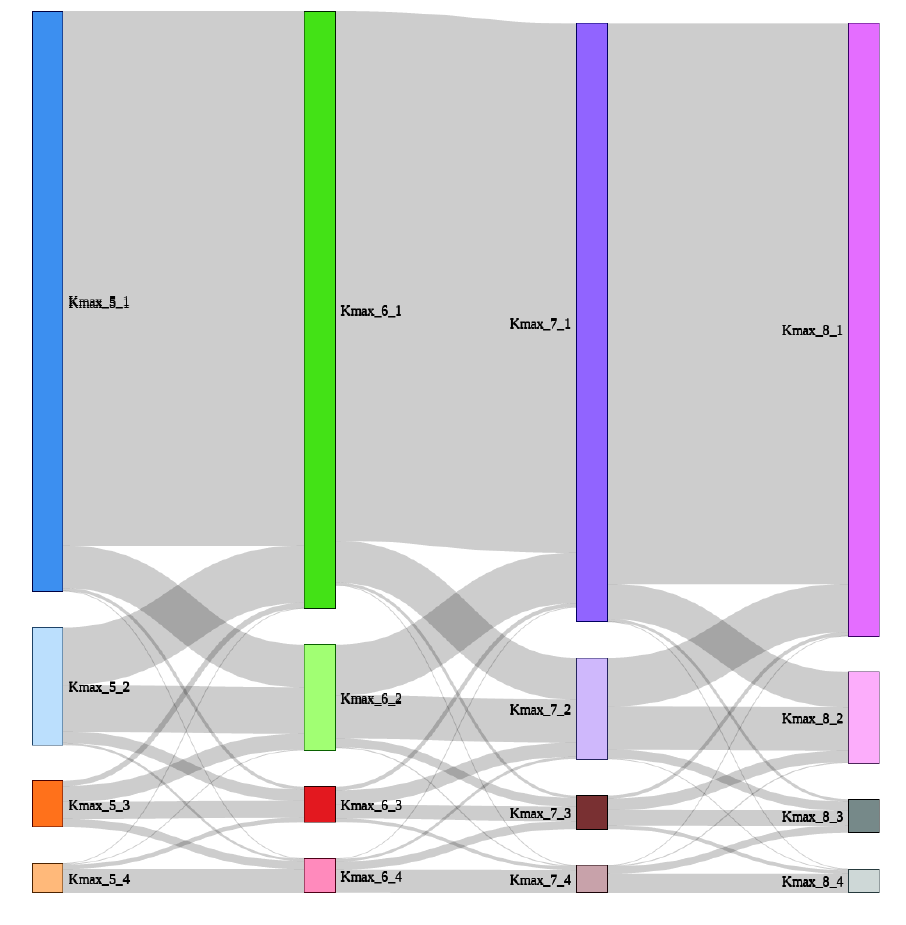}
\caption{\label{fig9} Migration of hashtags from various region of k-shells over 4 consecutive months from May 2012 to August 2012.}
\end{center}
\end{figure}

With the same intention of analyzing the stability of the core-periphery structure of the hashtag co-occurrence graph, we analyze the migration phenomenon of the hashtags across various regions (see fig~\ref{fig9}). The difference of this hashtag migration and the word migration discussed earlier are the following. More number of hashtags move from the core to the peripheral shells and vice-versa relative to the number of words; this is because hashtags are generally used to depict events in comparison to words and once the event loses its significance, the hashtags related to it become less popular and hence migrate to the peripheral shells. Nevertheless, there are some hashtags that remain persitent in the innermost shell over long periods of time; examples include \#amazing , \#android ,\#art , \#asian , \#awesome , \#baby , \#beach, \#beautiful, \#beauty, \#best ,\#bestoftheday ,\#birthday etc. These are more generic hashtags. Some of the hashtags which migrate from one shell to another are : \#
mothersday (moved from shell number 40 to shell number 2), \#prettylittleliars (moved from shell number 9 to 32) , \#thevoiceuk (moved from shell number 7 to 22). These mobile hashtags are mostly event-specific.
  \subsection{Hashtag repetition}
\begin{figure}[h]
\begin{center}
\includegraphics*[scale =0.3,angle=0]{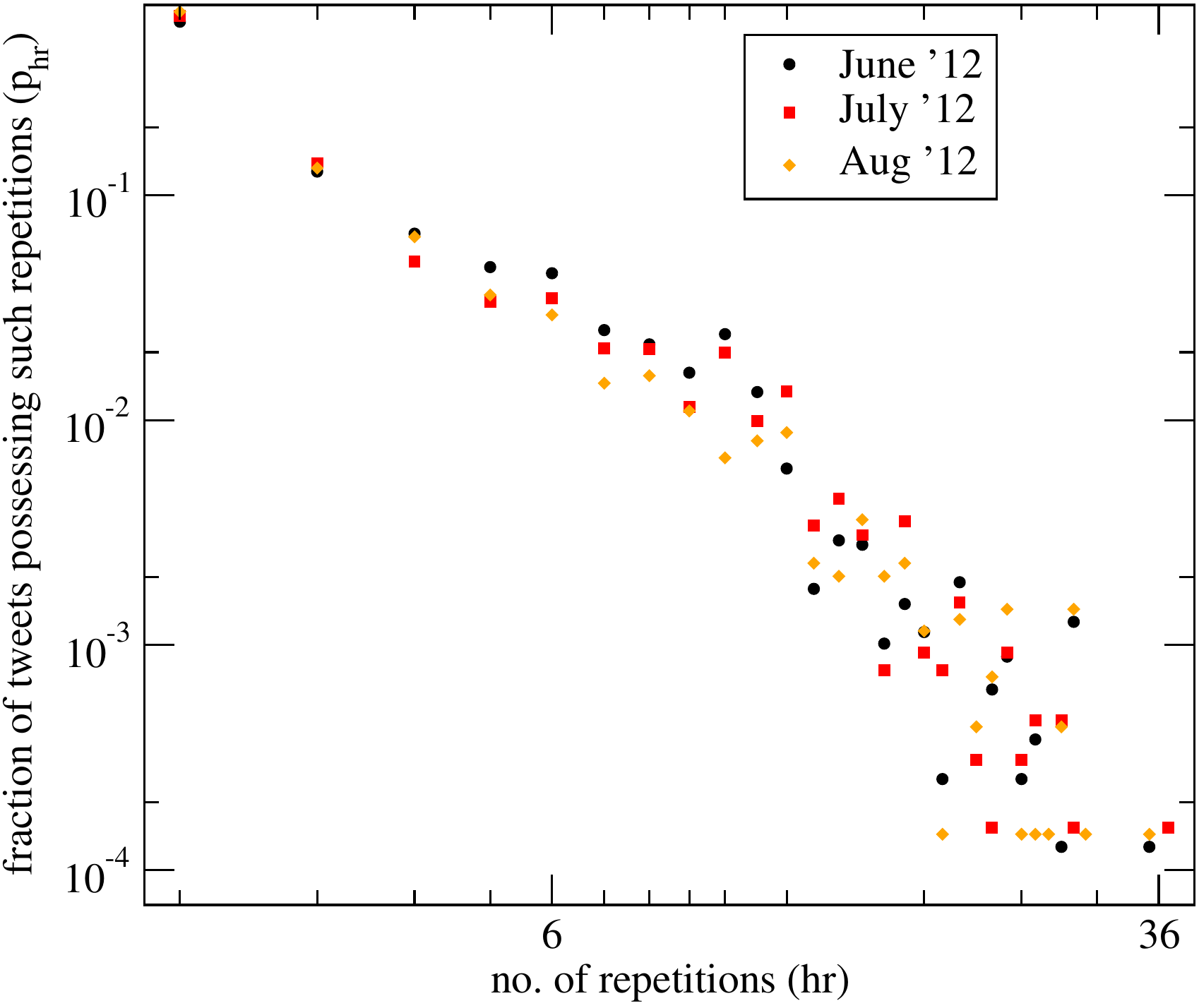}
\caption{\label{fig10} Distribution of repetition of hashtags for 3 representative months.}
\end{center}
\end{figure}

\begin{figure}[h]
\begin{center}
\includegraphics*[scale = 0.25,angle=0]{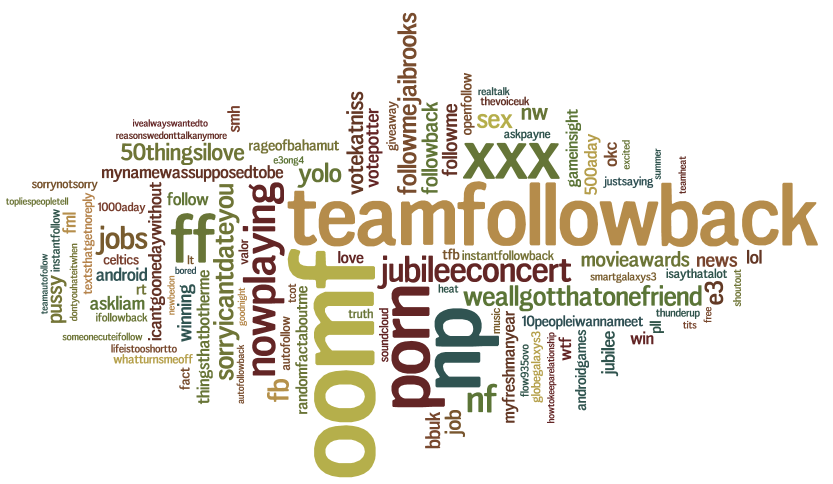}
\caption{\label{fig11} Hashtag cloud distribution for highly repetitive hashtags in June 2012 data.}
\end{center}
\end{figure}

In the previous section we have seen that in many cases multiple hashtags collocate in a single tweet. Another interesting linguistic phenomena that we observe in the tweets is the repetition of hashtags. People tend to repeat the hashtags when he/she is usually expressing a strong opinion on some issue/event. Also people use repetition of hashtags to express excitement or happiness. For example, we come across a tweet that contains only \#snow appearing in it mainly expressing a strong feeling of the user regarding possibly the current weather condition. In fig~\ref{fig10}, we show the degree of repetition of hashtags in tweets in the form of a probability distribution. The distribution follows power-law. In the tail of the distribution, we observe that there are instances of above 20 hashtag repetitions in tweets. This distribution is also consistent across all the different months. In fig~\ref{fig11}, we show the tagcloud of the most repetitive 
hashtags in June 2012 data where we clearly see that the `follow' hashtag, the slangs like np, oomf and hashtags with promotional activities are repeated widely. However, the hashtags that repeat large number of times across different months are the general expression words. For example, the hashtags like snow, heat, burn, omg, wow, rest, sleep etc. repeat more than 20 times in a tweet.

\subsection{Hashtag coalescing}
In etymology, we come across words that are formed from various other words sampled from the same or a different language. This linguistic phenomena of word coalescing is not new and we found many instances of word coalescing over the history of evolution of any language. For example, in English, `milkman' has been formed from `milk' and `man', walkman is the combination of `walk' and `man' with meaning of the words getting slightly modified due to coalesce. Similarly, `in so far' has become `insofar'. In today's world of brief expressions, chats etc. such merging phenomena in social media are far more prevalent than in standard texts and language. Further, such mergings happen at very short timescales compared to years/centuries in case of languages. In this subsection, we analyze the coalescing phenomenon of the hashtags i.e., how new hashtags are born from the merger of more than one hashtags. For example, \#peopleschoice and \#awards together form \#peoplechoiceawards. \#journals, \#
justinbieber, \#book form \#justinbieberjournalsbook ; \#mtvsports and \#justinbieber make \#mtvsportsjustinbieber; \#oregonbelievemoviemeetup is formed by \#oregon, \#believemovie and \#meetup; \#educational, \#ipad, \#apps together form \#educationalipadapps etc.

To identify the hashtag merging phenomena in Twitter, we take each month's 10000 most frequent hashtags and check if combination of these hashtags make a complete hashtag in the tweets posted after the time period in which these hashtags appeared. This does not result in all hashtag mergings but still we are able to identify a considerable number of such mergings. Merging of two hashtags are more frequent whereas merging of more than two hashtags also exist. For example, \#dontthinkaboutyouthatmuch consists of \#dont, \#think, \#about, \#you, \#that, \#much; \#takemeoutthegossip is formed by \#takemeout, \#the and \#gossip etc. These hashtags are commonly known as Twitter idioms.

\begin{figure}[h]
\begin{center}
\includegraphics*[width=1\columnwidth,angle=0]{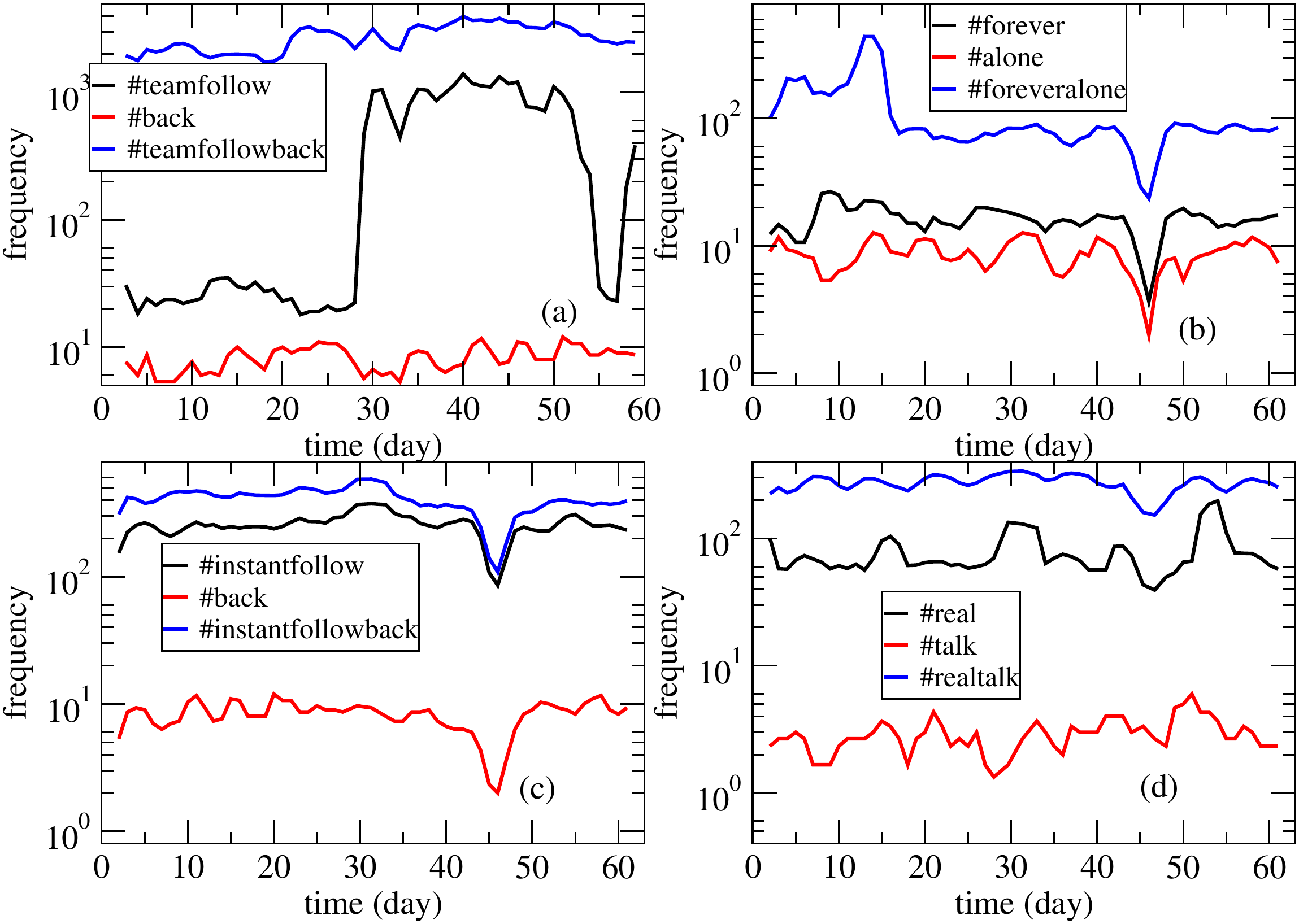}
\caption{\label{fig15} Frequency variation of the merged hashtag and its constituent hashtags over the days. The graphs are smoothed by taking moving window averages for better visualization.}
\end{center}
\end{figure}

In fig~\ref{fig15}, we show some examples of merged hashtags and their constituent hashtags from the point of merging. In this figure, we pick only those examples where the frequency of the merged hashtag is more than that of the constituent hashtags. 

\begin{figure}[h]
\begin{center}
\includegraphics*[scale = 0.27,angle=0]{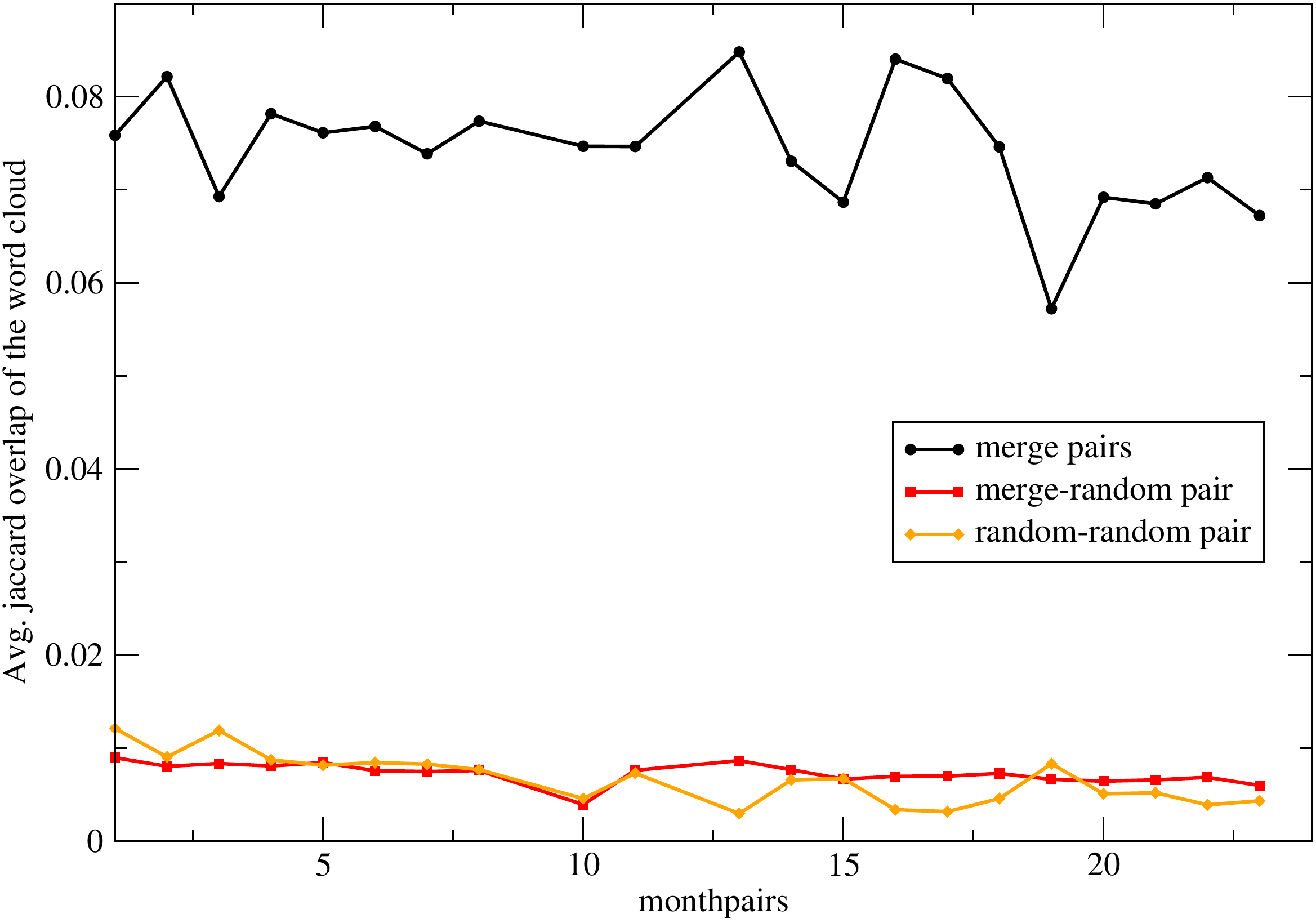}
\caption{\label{fig14} Jaccard Coefficients of the word cloud of merging hashtags, one merging and one random hashtag and both random hashtags of consecutive month pairs. The x-axis tick labels 1, 2 ...means January-February 2012, February-March 2012, ..respectively. }
\end{center}
\end{figure}

Next, we attempt to investigate a possible reason for such mergings. Why do two or more hashtags merge to form a separate hashtag? To identify the cause, we devise an experiment as follows. We consider all the hashtag mergings in which two hashtags get merged into one at a later timepoint. Note that our examples contain only those cases where the merged hashtag has a frequency higher than that of the constituents. For all such pairs of hashtags, we compute the Jaccard overlap of the word cloud around these hashtags. We then compute and plot the average overlap for all such hashtag pairs across consecutive months. A word cloud around a hashtag refers to all the words that the hashtag co-occurs with across all different tweets where the hashtag is present. For example, let's say \#A and \#B are getting merged to form \#AB. All the words of the tweets in which \#A appears, form the set word\_cloud (A). Similarly, for \#B, we consider word\_cloud (B). Now we compute the Jaccard index of the two sets of words word\_cloud (A) and word\_cloud (B). We observe a very high overlap between the word clouds around the hashtag 
pair. As these hashtags share similar informations, there seems to be social pressure on them to get merged. We validate our hypothesis across all the consecutive monthpairs (see fig~\ref{fig14}) and observe that it holds for all of them. To further strengthen the hypothesis, we perform a control experiment to check whether the overlap we achieved is better than the random case. For this purpose, we choose equal no. of i) random hashtag pairs across consecutive months ii) one hashtag among the merge-pairs and another completely random from the data and observe the avg. jaccard overlap among their word cloud. In both cases, we observe that the overlap is very low (at least 10 times lesser).
\section{Conclusions and future works}
In this paper, we study the sociolinguistic aspects of Twitter at a large time scale. To the best of our knowledge, this is the first comprehensive study on the evolution of the different sociolinguistic aspects of this OSN. We performed the analysis both on the word level as well as on the hashtags since it is perhaps the most important linguistic unit. We observe that it is inappropriate to claim that Twitter is more (in)formal because while on one hand, we see that OOV words are decreasing over time on the other hand it is clear that whitespace usage is getting reduced with a huge prevalence of running texts. We also observed that Twitter texts follow Zipf's law like natural language and has a strong core-periphery structure with words in the cores hardly migrating over time.

We perform similar linguistic studies on hashtags as we did on the words and observe that both the core-periphery, Zipf's law and other linguistic quantities show similar behavior. We also observe the hashtag repetition and hashtag coalescing phenomena and observe that there are sound reasons for the same. A remarkable observation is that the frequency of the coalesced hashtag is far greater than individual subparts. 

In future, we wish to study the hashtag coalescing, repetitions in more detail with more quantitative analysis and reasonings possibly tying them to certain evolutionary aspects of language. In this study, we have restricted ourselves to English tweets only, however in future, we wish to perform cross-linguistic study on Twitter texts. One interesting direction could be to understand how people with one language background can be benefited from other language discussions. Tweets are not always in one single language. They are code-mixed (two or more languages mixed in the text). This code mixing is very common phenomena in all multilingual societies. Not only Twitter, this is prevalent in other interactive social media platforms like Facebook, WhatsApp etc. The code-mixed words in the texts will be treated as OOVs and even without completely understanding the meaning of the whole tweet/text, one can possibly guess the meaning of the tweet texts by the clue OOV words in the text. This could help improving various language learning apps like duolingo\footnote{https://www.duolingo.com/} etc.
\section{acknowledgments}
The authors would like to thank Prof. Chris Biemann, TU Darmstadt for providing them with a historical Twitter 1\%
random sample data. This work has been supported by Microsoft Corporation and Microsoft Research India under the
Microsoft India PhD fellowship Award.
\bibliographystyle{IEEEtran}
\bibliography{ref}

\end{document}